\documentclass[pdflatex,sn-mathphys-num]{sn-jnl}

\usepackage{graphicx}%
\usepackage{multirow}%
\usepackage{amsmath,amssymb,amsfonts}%
\usepackage{amsthm}%
\usepackage{mathrsfs}%
\usepackage[title]{appendix}%
\usepackage{xcolor}%
\usepackage{textcomp}%
\usepackage{manyfoot}%
\usepackage{booktabs}%
\usepackage{algorithm}%
\usepackage{algorithmicx}%
\usepackage{algpseudocode}%
\usepackage{listings}%

\newcommand\B{{\scriptscriptstyle B}}
\newcommand\cv{c_{\scriptscriptstyle V}}
\newcommand\eqn[1]{eq.\ (\ref{#1})}
\newcommand\etc{{\sl etc.\/}}
\newcommand\fgn[1]{Figure \ref{fg:#1}}
\newcommand\gnew{G_{\scriptscriptstyle N}}
\newcommand\ie{{\sl i.\ e.\/}}
\newcommand\K{{\scriptscriptstyle K}}
\newcommand\mcons{\rotatebox[origin=c]{180}{m}}
\newcommand{\msol}{M_\odot}
\newcommand\mub{\mu_\B}
\newcommand\mui{\mu_{\scriptscriptstyle I}}
\newcommand\muq{\mu_\Q}
\newcommand\mus{\mu_\St}
\newcommand\Pc{{\cal P}}
\newcommand\pic{$\pi$C}
\newcommand\ppbar{\langle\overline\psi\psi\rangle}
\newcommand\Q{{\scriptscriptstyle Q}}
\newcommand\Sc{{\cal S}}
\newcommand\St{{\scriptscriptstyle S}}
\newcommand\tco{T_{co}}

\begin{document}

\title[Phases of QCD]{The phases of QCD reached in terrestrial and cosmic colliders}

\author*{\fnm{Sourendu} \sur{Gupta}}\email{sgupta@theory.tifr.res.in}

\affil{\orgdiv{International Center for Theoretical Sciences},
 \orgname{Tata Institute of Fundamental Research}, 
 \orgaddress{\street{Survey 151 Shivakote, Hesarghatta Hobli}, 
 \city{Bengaluru North}, \postcode{560089}, \country{India}}}

\abstract{
We review the current state of knowledge of the phase diagram of
QCD through lattice, effective field theories, and chiral models.
Several sections through the three dimensional phase diagram are known
for $N_f=2+1$ with good precision.  Due to technical advances in lattice
techniques over the last decade or so, new aspects of the phase diagram
can now be explored. We review current lattice results. The newly
acquired knowledge can be used to reconstruct the full phase diagram
for physical QCD, \ie, $N_f=1+1+1$. We remark on the computations which
would help understand this better, and what the current constraints are
on matter in neutron star cores. We also remark on the physics of the
chiral transition and neutron stars in the 't Hooft large $N_c$ limit.
}

\keywords{QCD, phase diagram, heavy-ion collisions, neutron stars}

\maketitle

\section{The space of the phases of QCD}\label{sec:one}

Although field theoretic foundations for the computation of strong
interaction thermodynamics were established long back, there are many
challenges in its use. Not only is computing power an issue, but there
are also some foundational issues with its use, for example the fermion
sign problem.  One aim of this work is to review some recent methods
and results from lattice QCD computations. A second aim is to use this
information and basic thermodynamic arguments to try to constrain the
phase diagram of $N_f=1+1+1$ QCD. This allows us to address the connection
between heavy-ion collision (HIC) experiments and the properties of
neutron stars (NS), and where theoretical methods help to connect the two.

\begin{figure}
\centering
\includegraphics{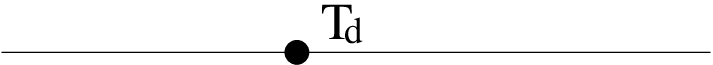}
\caption{The thermodynamics of a gluon gas has two dimensional Gibbs
space labelled by $E$ and $S$. The phase diagram, shown here, is one
dimensional and labelled by $T$. If there are confined and deconfined
phases, then Gibbs' phase rule indicates that the phase diagram has a
single first order transition, labeled here as $T_d$.}
\label{fg:onedim}
\end{figure}

When a system comes to equilibrium all the quantities that describe its
changeable character are gone, and only the conserved quantities remain.
For strongly interacting matter these are the energy ($E$), baryon number
($B$), and its electrical charge ($Q$). So the thermodynamic description
of this matter involves these extensive quantities, along with the entropy
($S$). This is true whether we consider such matter in the early universe,
in extreme astrophysical conditions, or in the lab. Of course, equilibrium
does not mean static. At microscopic scales the conserved quantities
may move from one constituent to another, but in the coarse grained
description that we call thermodynamics, this is irrelevant.

In this four dimensional Gibbs space, states in thermodynamic equilibrium
are described by the convex entropy surface $S(E,B,Q)$. A different
description of the same surface is given by the function $E(S,B,Q)$. The
slopes of this energy surface, \ie, the partial derivatives of this
function $E$ with respect to its arguments, defines the thermodynamic
intensive quantities $T$, $\mub$ and $\muq$. An equivalent description is
obtained by working in the space of phases of matter, which is labelled
by these intensive quantities.

From $E$ one defines, through successive Legendre transformations,
the thermodynamic potential, $\Omega(T,\mub,\muq)$. In each distinct
phase of matter there is only one continuous and convex function
$\Omega$. However, when there are multiple phases, each may have a
distinct function $\Omega$. A thermodynamic system always settles into
the phase with the lowest $\Omega$. We argue next that the phase diagram
can be thought of as a plot of the discontinuities and singularities of
the thermodynamic potential as a function of the intensive quantities.

At some points it is possible for two phases, say $a$ and $b$, to
have equal values of the potential. Then we say that the two phases can
coexist, or in the old language formalized by Ehrenfest, that there is
a first order transition between them. Obviously, with three variables,
one equation $\Omega_a (T, \mub, \muq) = \Omega_b (T, \mub, \muq)$,
in general gives two-dimensional surfaces of solutions. So there are
surfaces of phase coexistence.  Along these surfaces the change in entropy
from one phase to another is called the latent heat. When this vanishes
the functions $\Omega_a$ and $\Omega_b$ are no longer distinguishable,
and the coexistence surface has come to an end.  It can be proved that
both $\Omega_a$ and $\Omega_b$ have a singularity when the latent heat
vanishes \cite{Buckingham:1972}. This is called a critical point, or in
the older language of Ehrenfest, a second order transition. Since two
dimensional surfaces end along curves, these edges are critical lines
in the phase diagram.

The only other way for a surface of coexistence to end is by meeting
another such surface. Again this happens along a line. On this line three
phases coexist. If one of the surfaces also happens to end in a critical
line, then the intersection of the critical line with the three phase
coexistence line gives rise to a tricritical point. Arguments such as
these are Gibbs' phase rules.

The simplest thermodynamic systems are pure gauge theories. Since the only
conserved quantity is $E$, the Gibbs space is two dimensional, with
extensive quantities $E$ and $S$.  The space of phases is one dimensional
and has the coordinate $T$, where $dE=TdS$. There may be a single phase
in a pure gauge theory, as in the case of pure gauge electromagnetism
(EM), \ie, a photon gas. There are no phase transitions. However,
non-Abelian pure gauge theories can have two phases: either a confined
phase or the non-Abelian analogue of the photon gas. Hence there can be
two thermodynamic potentials. As a result, Gibbs' phase rule indicates
that there can be a single first order phase transition between these
two phases. This is called the deconfinement phase transition.

For SU($N_c$) gauge theory with $N_c=2$ the deconfinement phase
transition is of second order \cite{Yaffe:1982qf, Polonyi:1982wz}. This
violates Gibbs' phase rule. One notes however, that the order parameter
for deconfinement is related to the free energy of a quark. A quark
coupled to SU(2) gauge theory has an extended symmetry compared to any
other number of colours, $N_c$. This prompts us to examine the phase
diagram of pure gauge theories as a function of both $T$ and the theory
parameter $N_c$. In this plane, there is a line of first order phase
transitions for $N_c>2$ \cite{Datta:2009jn} which ends in a critical
point at $N_c=2$. This is completely compatible with Gibbs' phase
rule. In fact we also have the ``prediction'' that a photon gas has no
phase transition.  What we learn from this is that apparent violations of
the Gibbs' phase rule are possible, and are due to enhanced symmetries.
In that case one has to investigate a theory parameter which breaks the
symmetry generically. It is interesting that thermodynamics predicts
these topological features of the phase diagram of non-Abelian pure
gauge theories without any reference to the microscopic physics.

Matter at high temperature or density in nature is not made solely of
strongly interacting particles. Photons and neutrinos in the early
universe may be an important part of the mixture. Stability over
macroscopic distances also requires local charge neutrality. Departures
from this condition will result in internal currents in matter
which eventually restore the neutrality and bring the system back to
equilibrium. In such mixed systems consideration of the conserved lepton
number may be needed.

In HICs the huge separation of time scale between the strong and weak
interactions means that strangeness may be considered to be conserved
in the strongly interacting fireball. A full description of the
corresponding thermodynamics would then require us to take into account
the net strangeness, and the conjugate intensive variable, namely the
strangeness chemical potential $\mus$. Such a four dimensional phase
diagram may also be relevant in neutron stars (NSs) where stability
against beta-decay may be achieved with a large strangeness density
\cite{Alford:1999pa}. However, studies in this full four dimensional
space are very limited, and we will consider the slice with $\mus=0$,
since this seems to suffice for HICs and many NSs.

This completes a list of the interesting dimensions of the phase
diagram of strongly interacting matter. There are also parameters in
the QCD Lagrangian, namely the quark masses, $m_u$, $m_d$ and $m_s$.
They are tuned non-perturbatively to reproduce known meson and baryon
masses. However, in order to understand the phase diagram, we have
to understand the effect of chiral symmetry of QCD when one or more
masses vanishes: this is chiral QCD. Then it is convenient to study
QCD by changing the quark masses. So the theoretical space of phases
contains three more dimensions (the quark masses) than the phase diagram.
For example, one particular section of the theory space ($m_\ell=m_u=m_d$,
$\mub=\muq=\mus=0$), namely the Columbia plot, has long been studied
for the insights it provides into the problem \cite{Brown:1990ev,
Cuteri:2021ikv}. One major result is that the topology of the phase
diagram is extremely insensitive to $m_s$, provided that the pion mass
is not ultralight (much lighter than the physical value) or ultraheavy
(much heavier than physical).

It is common to use the notation $N_f=3$ for the slice $m_u=m_d=m_s$,
and $N_f=2+1$ for $m_u=m_d<m_s$.  The breaking of $N_f=3$ to $N_f=2+1$
is accomplished by tuning the light and strange quark masses, $m_\ell$
and $m_s$, in order to reproduce the isospin averaged values of the meson
masses $m_\eta$, $m_\K$, and $m_\pi$. We will also discuss chiral QCD,
for $N_f=2+1$ this has $m_\pi=0$ but the other mesons have non-vanishing
mass. The physical theory is QCD with $N_f=1+1+1$.  Breaking $N_f=2+1$
to $N_f=1+1+1$ would entail tuning the difference between d and u
quark masses, $\Delta m$, in order to further reproduce the physical
value of $m^2_{\pi^0}/m^2_{\pi^\pm}$. Since part of the difference in
masses of charged hadrons and their uncharged isospin partners is due
to electromagnetic (EM) effects, this tuning is quite finicky. So it
is interesting to examine the topology of the phase diagrams with changing
$\Delta m$ to see whether it is sensitive to this parameter.

In various situations of interest, either in NSs or in HIC experiments,
strongly interacting matter may be immersed in large external magnetic
fields, $B$. In such cases the influence of magnetic field on the
phase diagram may be of interest. The energy density of strongly
interacting matter is controlled by a scale close to the pion mass,
$m_\pi$. Clearly $B\simeq{\cal O}(m_\pi^2)$ is necessary for magnetic
fields to influence strong thermodynamics. Magnetic fields as large as
these could be produced momentarily in the very early stages of HICs,
but there is no clear evidence that such large fields persist until matter
is thermalized. In astrophysics, the largest magnetic fields, as much as
$10^{15}$ Gauss, are found in magnetars. However, $10^{15}\,{\rm Gauss} =
10^{-3} m_\pi^2$, so the change in the phase diagram due to such magnetic
fields is expected to be minimal. In this review we will ignore this
effect (see \cite{Endrodi:2024cqn} for a recent review of developments).

Other theory parameters can also be tuned. QCD-like theories
with different number of colours, $N_c$, have been studied \cite{
McLerran:2007qj, Datta:2010sq, Lucini:2012gg, McLerran:2018hbz}. As
long as the conserved global quantities are the same, their study
can throw light on QCD.  Recently interesting new aspects of QCD
phases have been found when $N_f$ and $N_c$ are varied simultaneously
\cite{Cherman:2017tey}.  Also of deep theoretical interest is the question
of how the broken axial $U(1)$ symmetry of QCD affects thermodynamics
and the phase diagram \cite{Aoki:2020noz, Gavai:2024mcj, Fodor:2025yuj}.
Among lattice techniques, there have been advances \cite{Mondal:2021jxk,
Mitra:2022vtf} in the method of differential reweighting
\cite{Allton:2002zi} by a Taylor expansion of the quark determinant
inside the path integral.  Discussion of these questions lie outside the
scope of this review. The technique of Functional Renormalization Group
\cite{Polonyi:2001se, Metzner:2011cw, Dupuis:2020fhh} has recently been
used for QCD \cite{Gao:2020fbl}. This subject requires a separate review.

\section{The $T$-$\mub$ plane}\label{sec:two}

The part of the phase diagram of strongly interacting matter that can
be probed in relativistic heavy-ion collisions has large variations
in $T$ and $\mub$ and relatively small variations in $\muq$. Lattice
computations suffer from a sign problem when $\mub\ne0$ is introduced into
the QCD action, but are able to give accurate and precise information for
$\mub=0$. Taylor expansions of thermodynamic variables can then be used
to extrapolate to finite $\mub$. We examine the situation with $\mub=0$
before moving on to the rest of the phase diagram. In these arguments
the relevant order parameter is the quark condensate $\Sc=\ppbar$. Its
value distinguishes between the hadron phase, where $|\Sc|$ is large,
and the quark phase, where it is small, and vanishing in chiral QCD.

\subsection{Lattice results}

\begin{figure}
\centering
\includegraphics[scale=0.5]{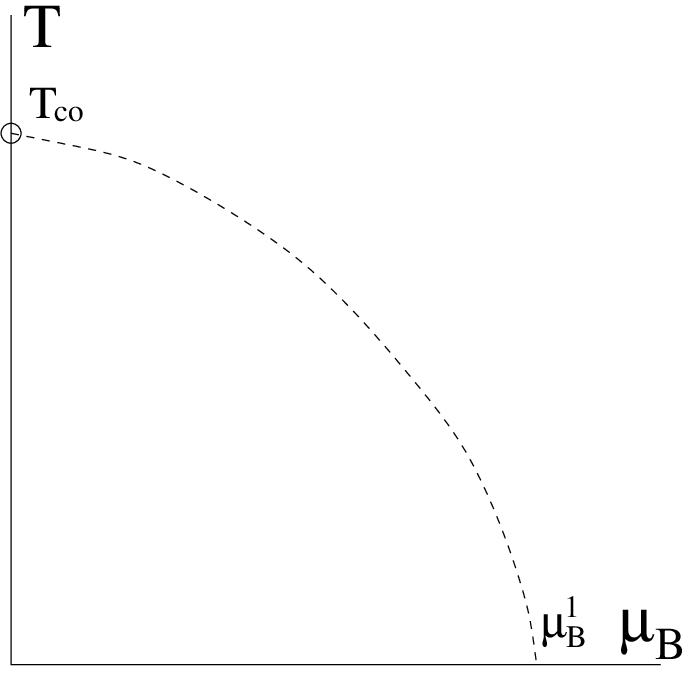} \hfill
\includegraphics[scale=0.4]{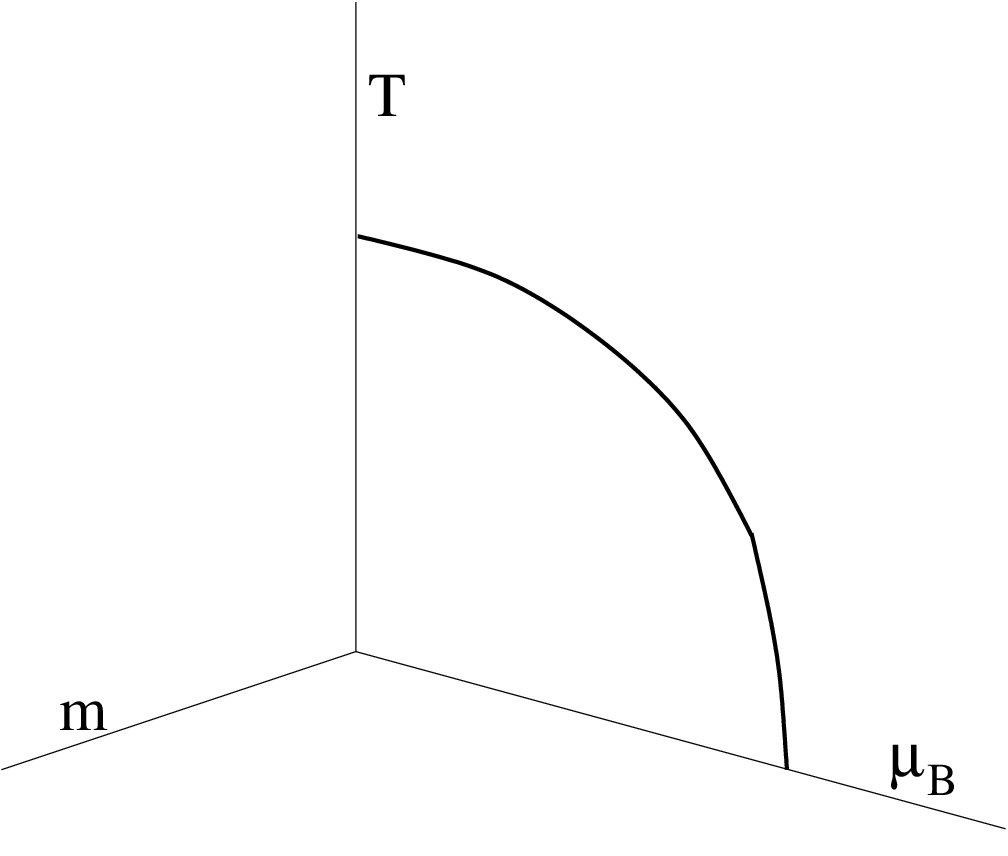}
\caption{The most parsimonious phase diagram for QCD with $N_f=2+1$
based on current lattice results and experiments. At fixed and
non-vanishing pion mass it would yield only a crossover (panel on
the left).  An estimate of $\mub^1$, where the crossover line crosses
the axis is given in the text.  This is a shadow of a critical line in
the chiral limit, with a measured curvature (panel on the right). This
critical line bounds a surface in chiral QCD at which different signs
of $\Sc$ coexist.}
\label{fg:classic}
\end{figure}

It has been known for a long time that there is no phase transition in
QCD with two light flavours and one heavy flavour (strange) of quarks
\cite{Brown:1990ev, Cuteri:2021ikv, Pisarski:1983ms, Aoki:2006we}. Instead
of a discontinuous or singular behaviour of thermodynamics at a phase
transition, one has a continuous change of behaviour. However it is
abrupt enough to give interesting experimental signals. This is a
cross over--- the speed of sound may go through a non-zero minimum at one
temperature, a susceptibility can have a non-singular peak at another
temperature $\tco$, {\etc} The temperature interval where such extrema
occur has a non-zero width $\Delta T$. Different maxima and minima can
occur anywhere in this interval.

In chiral QCD then there is second order transition
\cite{Pisarski:1983ms}. Since $\mub$ does not affect the symmetry which
is broken or restored across $T_c$, this critical point develops into
a critical line, which can be parametrized at small $\mub$ as
\begin{equation}
  T_c(\mub) = T_c\left[ 1 - \kappa_2\left(\frac{\mub}{T_c}\right)^2
		- \kappa_4\left(\frac{\mub}{T_c}\right)^4 + \cdots \right],
\label{curvs}
\end{equation}
where we used the notation $T_c$ for the more cumbersome $T_c(\mu=0)$. The
coefficients $\kappa_2$ and $\kappa_4$ are called curvature coefficients.
Flipping quarks with antiquarks is equivalent to $\mub\leftrightarrow
-\mub$, and the flip does not change any physics. That's why the expansion
in \eqn{curvs} contains only even powers of $\mub$.

Many modern day computations of thermodynamics are performed with improved
staggered quarks with two degenerate light flavours and a somewhat heavier
flavour (strange).  The improvement consists of control over lattice
spacing effects, so that continuum extrapolations are straightforward.
Bare quark masses are tuned to reproduce the mean masses of isospin
multiplets of the pseudoscalar SU(3) flavour octet mesons properly.
Two sets of extensive computations for the phase diagram can be compared.
These are results at finite $T$ and vanishing charges and external fields
\begin{align}
\tco=156.5\pm1.5 {\rm\ MeV\/}, &
\quad\kappa_2^\B=0.015(4),
\quad\kappa_2^\Q=0.027(4), \nonumber\\&
\quad\kappa_4^\B=-0.001(3),
\quad\kappa_4^\Q=0.004(5),
\label{curv}
\end{align}
from \cite{HotQCD:2018pds}.  Here $\kappa_{2,4}^\Q$ are curvatures in
the $\muq$ direction.  These results are statistically consistent with
those from \cite{Borsanyi:2020fev},
\begin{align}
\tco = 158.0\pm0.6 {\rm\ MeV\/}, &
\quad\Delta T = 15 \pm 1 {\rm\ MeV\/}, \nonumber\\&
\quad\kappa_2=0.0153(18),
\quad\kappa_4=0.00032(67)
\label{bors}
\end{align}
It is instructive to note that although a particular measurement of $\tco$
can be made more and more precise through more statistics and by careful
control of various systematic errors, the width of the cross over region,
$\Delta T$ is expected to remain non-vanishing. The most parsimonious phase
diagram with all these inputs is shown in \fgn{classic}. It is compatible
with old expectations that followed from \cite{Pisarski:1983ms}.

The lattice determinations of $\tco$ for varying quark masses have been
extrapolated to chiral QCD using the scaling exponents for the flavour
symmetry group SU$_L$(2)$\times$SU$_R$(2) $\simeq$ O(4). This gives
\cite{HotQCD:2019xnw}
\begin{equation}
T_c=132^{+3}_{-6} {\rm\ MeV\/}.
\label{chiral}
\end{equation}
In all the results quoted here, lattice spacing effects are under good
control. Reliable simulations are now available at quark masses which
are not only realistic, but also for masses lower than physical. This
improves the reliability of the extrapolation to chiral QCD.

A new generation of simulations of thermodynamics with improved Wilson
quarks has started \cite{Brandt:2019ksy, Bresciani:2023zyg}, but the pion
masses which are accessible are generally higher. Still, one can look
forward to better quantification of the effects of lattice spacing by a
comparison of two different methods of putting quarks on the lattice. Most
simulations use a spatial lattice volume which is pretty large. However,
in future one should look forward to a proper finite size scaling study
of the cross over, especially in order to control the extrapolation to
chiral QCD as the quark mass is decreased below the physical mass

\subsection{Effective field theory}

Chiral symmetry has been used in various ways before. Famously, the
NJL model has been used to understand the effects of this symmetry,
and has been found to be numerically fairly accurate for several kinds
of quantities \cite{Klevansky:1992qe}. More importantly, at $T=0$
an effective field theory (EFT) called chiral perturbation theory
($\chi$PT) is known to work extremely well \cite{Gasser:1983yg} (see
also \cite{Shuryak:1992pi}) for low energy hadron physics. Its success
has prompted lattice computations at $T=0$ to concentrate on precision
measurements of amplitudes which constrain the parameters of $\chi$PT.
the EFT is then used to compute amplitudes which are hard to access
using lattice methods.  $\chi$PT has also been extrapolated to finite
temperature \cite{Gasser:1987ah, Gerber:1988tt}, albeit with less success.

Recently the NJL model has been generalized to a thermal EFT by adding
all possible terms which are allowed by symmetry (apart from flavour
symmetry one also incorporates the broken Lorentz invariance for $T>0$),
and organized by the scaling dimension ($D$) of operators. Keeping all
the operators up to $D=6$, one can find a good description of lattice
measurements of static properties of pions using only three pieces of
data from lattice computations to fix the low-energy constants (LECs)
\cite{Gupta:2017gbs}.  Hartree-Fock treatment of this theory one
can recover a finite temperature analogue of $\chi$PT.

Interestingly if one fits only pion properties at $T\ge0$ for $N_f=2+1$
QCD then the phase diagram can be recovered quantitatively from the
EFT \cite{Gupta:2025kro}. One finds $\tco$, $\kappa_2$ and $\kappa_4$ in
excellent agreement with \eqn{curv} and \eqn{bors} within 68\% confidence
limits (CL). The relatively large errors in lattice measurements of the
curvatures is the weakest quantitative test of the lattice at present.
halving the errors in the lattice measurement of $\kappa_2$ would test
the accuracy of the EFT quite stringently.  In the limit of chiral QCD,
$T_c$ from the EFT is also in agreement with \cite{HotQCD:2019xnw}.

The EFT can also give results which are currently impossible to
obtain from the lattice. One such is the pole mass of pions at finite
temperature. The difference between screening and pole masses is a direct
consequence of thermal effects, and is captured in an EFT through the
fact that broken Lorentz symmetry (in the Euclidean theory) allows the
kinetic term in the pion Lagrangian to become
\begin{equation}
  \frac12(\partial_\mu\pi)^2\xrightarrow{T>0}\frac12(\partial_t\pi)^2
    + \frac12 u_\pi^2(\nabla\pi)^2.
\label{upi}
\end{equation}
In chiral QCD the pole mass, $m_\pi$, vanishes for $T<T_c$ and $u_\pi$
drops to zero at $T_c$ with a critical exponent \cite{Son:2002ci} Using
the pion propagator from \eqn{upi} one finds the Debye screening mass
$m_\pi^D=m_\pi/u_\pi$.  The EFT for $N_f=2+1$ predicts that when the
$T=0$ pion has mass of 140 MeV, the pole mass drops to about 110 MeV at
$0.85\tco$ and to about 100 MeV at $\tco$. Lattice measurements show
that $m_\pi^D$ increases faster than linearly with $T$ for $T\simeq
T_c$. This is due to the fact that the physical pion mass is close
enough to vanishing that $u_\pi$ drops near $\tco$ although it does
not vanish.  Due to this behaviour of $u_\pi$, the pressure, $P/T^4$,
also shows a rapid rise near $\tco$. However, it seems that NLO terms in
thermal $\chi$PT are likely to be needed in order to give a quantitative
description of the pressure below $\tco$.

\begin{figure}
\centering
\includegraphics[scale=0.4]{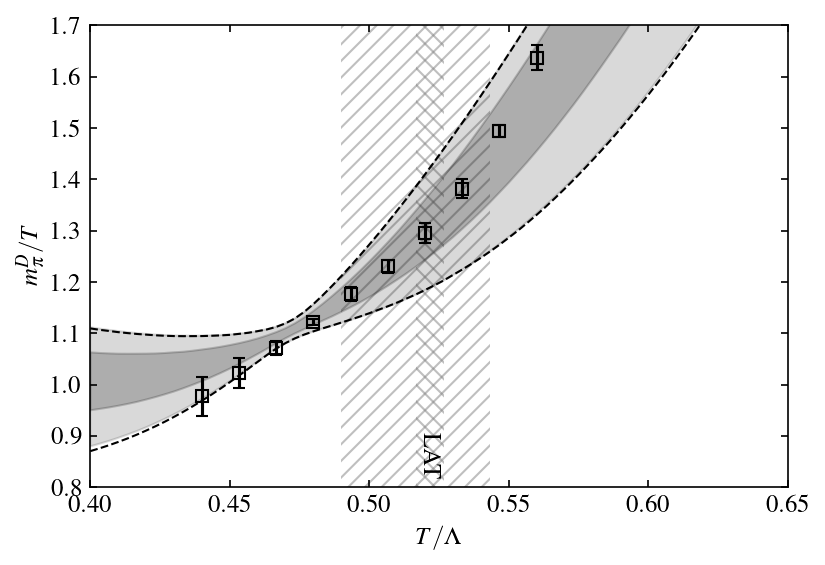} \hfill
\includegraphics[scale=0.4]{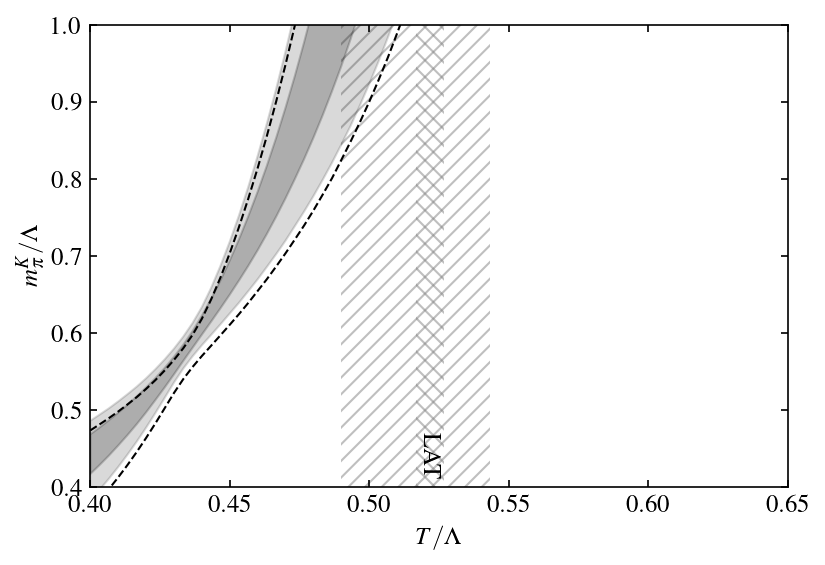}
\caption{The panel on the left shows that the screening mass of the pion,
$m_\pi^D$, is found to be continuous across $T_c$ in lattice computations.
The data points show continuum extrapolated results from the lattice. This
is reproduced well by the EFT (dark shading for the 68\% error band, light
shading for the 95\% error band), which has a UV cutoff $\Lambda$. This
particular fit uses $\Lambda=300$ MeV, but equally good fits are obtained
using $\Lambda=450$ MeV. The vertical hatched band shows $\tco$ (the narrow
band is from the lattice, the wider band from the EFT). The panel on the
right shows the kinetic mass, $m_\pi^K$. Since it becomes comparable to
$\Lambda$ at about $\tco$, pions are no longer the low-energy excitations
of QCD.}
\label{fg:softhard}
\end{figure}

The Minkowski version of the model Lagrangian in \eqn{upi} then shows
that the energy of a pion of 3-momentum $p$ is
\begin{equation}
   E_p = m_\pi + \frac12\,\frac{p^2}{m_\pi^K},
\label{mkin}
\end{equation}
where the kinetic mass $m_\pi^K = m_\pi/u_\pi^2$. Since $u_\pi$ decreases
towards $\tco$, one sees that $m_\pi^K$ increases \cite{Gupta:2020zqo}. As
a result, pion gains energy more slowly with increasing $p$ close to
$\tco$ than it would have at a lower temperature. This would close off
certain inelastic reactions as one approaches $\tco$.

Through this mechanism the EFT helps us to untangle the specific physics
of the QCD crossover. The continuity of screening masses across $\tco$
has been known for a long time. It is reproduced well by the EFT, as shown
in \fgn{softhard}. This means that appropriate probes can pick out pions
in hot matter. However, the kinetic mass of the pion rises rapidly, showing
that as far as dynamical processes are concerned, the pion can no longer
be considered a light collective mode near $\tco$. Other fluctuations of
quarks must then be investigated. The presence of static pions and its
simultaneous absence from dynamics is one aspect of QCD that lattice
computations presumably contain, but which can be made explicit only after
Wick rotation. The EFT is an effective tool for this.

\subsection{The physical phase diagram}

\begin{figure}
\centering
\includegraphics[scale=0.25]{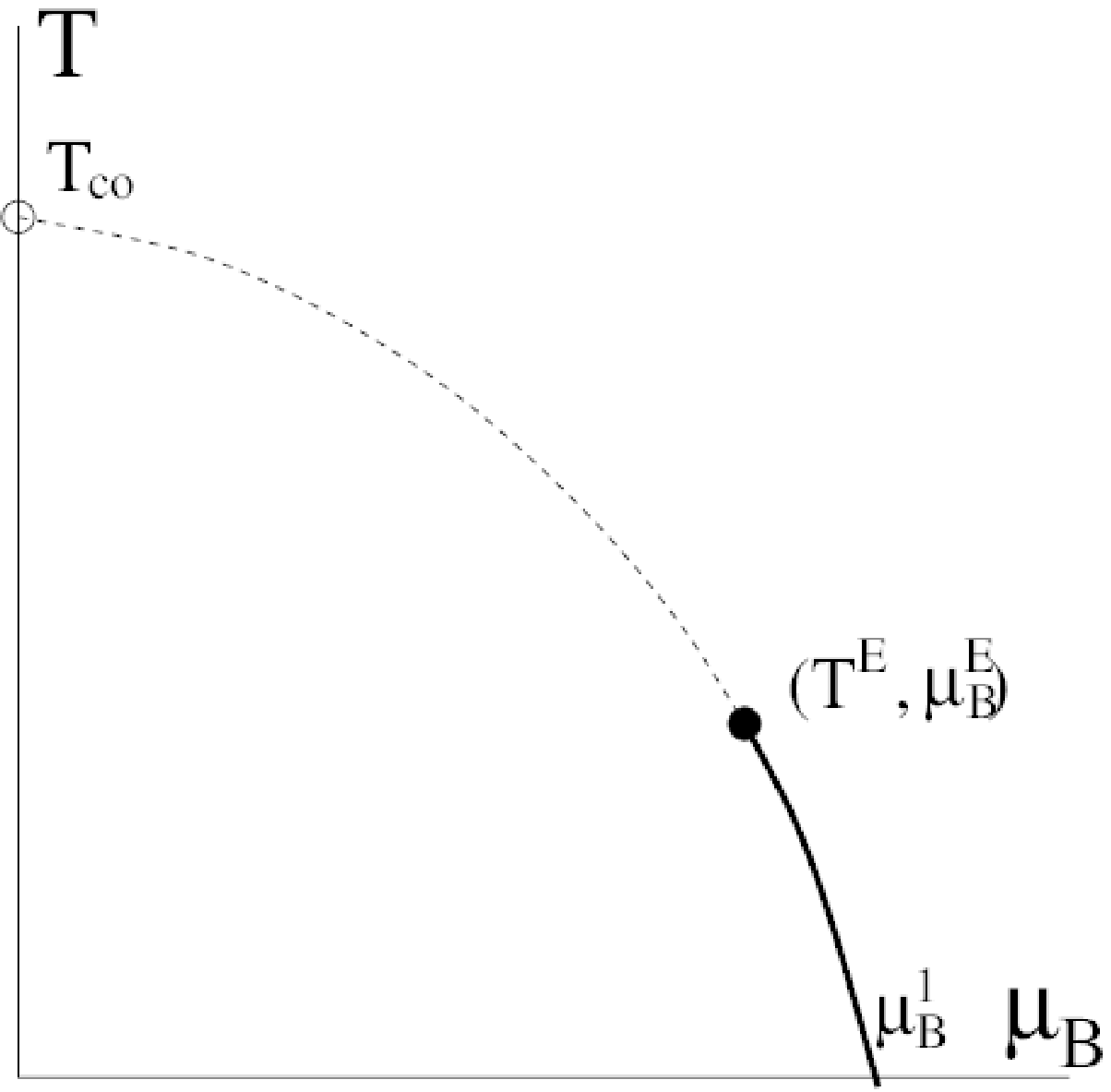} \hfill
\includegraphics[scale=0.3]{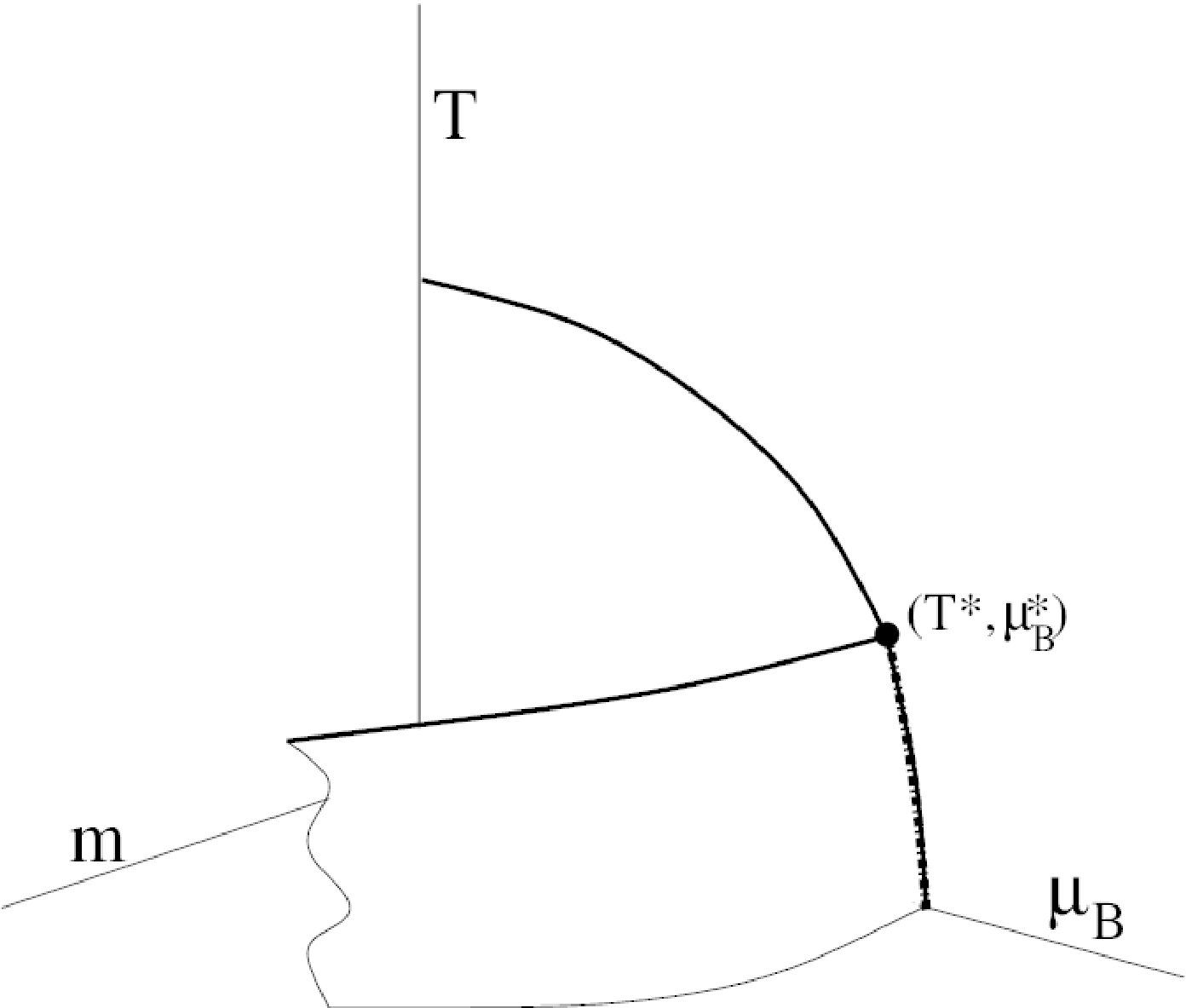}
\caption{The conjectured phase diagram for $N_f=2+1$ QCD with a finite
quark mass (panel on the left). The crossover at $\mu_\B=0$ develops
into a line of crossovers which turns into a line of phase coexistence (a
first order transition between hadron and quark phases) at the critical
point $(T^E,\mub^E)$. When extended in quark mass (panel on the right),
the critical point turns into a critical line, which is the boundary of
a surface of hadron-quark phase coexistence (part of it is cut away to
show the structure). This critical line meets the chiral critical line
at the tricritical point $(T^*,\mub^*)$, which is the end point of a
triple line in QCD.  The shape of this phase diagram is not expected to
change in $N_f=1+1+1$.  However, $\tco$, $\mub^1$, and $(T^E,\mub^E)$
could shift.}
\label{fg:critpt}
\end{figure}

The structure of the phase diagram of $N_f=2+1$ QCD with realistic
meson masses is conjectured on the basis of effective theories
\cite{Barducci:1993bh, Halasz:1998qr, Stephanov:1998dy}. Only the order
parameter $\Sc$ is involved in \fgn{critpt}. There is a critical line
in chiral QCD which stretches from $T_c$ at $\mub=0$ to the tricritical
point, $(T^*,\mub^*)$.  This critical line is the boundary of a surface
of first order transitions between $\Sc<0$ for $m>0$, and $\Sc>0$ for
$m<0$. The tricritical point is the end point of a triple line where
the three coexisting phases are the two chiral symmetry broken hadron
phases, and the third is the chiral symmetry restored quark phase with
$\Sc=0$. Although the phase diagram of \fgn{classic} is compatible with
all data till now, there are large portions of the phase diagram which
have not been studied and could yield evidence for the alternative
discussed here.

If first order transitions do occur at finite quark mass, then the
critical point, $(T^E,\mub^E)$, must lie on the cross over line starting
at $\tco$ with curvatures which are well determined. The earliest
lattice computations for finite $\mub$ which used Pad\'e approximants
found the end point \cite{Datta:2016ukp} at $\mub^E/T^E=1.85\pm0.04$
and $T^E/\tco=0.94\pm0.1$.  This lies squarely on the cross over curve
determined by the parameters in \eqn{bors} and \eqn{curv}. More recent
computations using multi-point Pad\'e \cite{Clarke:2024ugt} report
$T^E=105^{+8}_{-18}$ MeV and $\mub^E=422^{+80}_{-35}$ MeV. This is
also compatible with the shape of the cross over curve at $\mub=0$.
Interestingly, this work also reported a value of $\kappa_2$ at the
end point which is in agreement with that at $\mub=0$. This is possibly
the first analysis of lattice data which indicates that the low-order
expansion in \eqn{curvs} is justified for extrapolation to large $\mub$.

This observation is an useful estimate for a quantitative determination
of the coexistence surface between the hadron and quark phases. The
values in \eqn{curv} and \eqn{chiral} predict that for chiral QCD the
surface reaches $T=0$ for $\mub=1100\pm150$ MeV.  Using the same logic,
the first order line at physical meson masses ends at $\mub^1=1280\pm170$
MeV for $T=0$ using the values in \eqn{curv}. Using \eqn{bors} we find
that $\mub^1=1280\pm75$ MeV.

The phase diagram for realistic pion and kaon masses does not change
qualitatively between $N_f=2$ and $N_f=2+1$, although $\tco$ changes
in value as the strange quark mass is changed \cite{Karsch:2000kv}. For
physical QCD, \ie, with $N_f=1+1+1$ and realistic pion and kaon masses,
the phase diagram is not expected to change qualitatively. It is possible
however that $\tco$, the curvature coefficients of the cross over line,
the location of the end point, and $\mub^1$ may change somewhat. The only
investigation of $N_f=1+1$ QCD \cite{Gavai:2002fi} reported that no change
in $\tco$ was observed when changing the ratio $m_{\pi^0}^2/m_{\pi^\pm}^2$
between unity and 0.78 by changing $\Delta m$. It would be useful to
bring modern techniques and improved statistics to bear not only on $\tco$
but also the curvature coefficients in physical QCD.

If other order parameters are involved at such low $T$ and for such
$\mub$, for $N_f=1+1+1$ QCD, then the physical phase diagram may be more
complicated than what is shown here. However, at this time there is no
clear evidence for such complications. At very low temperatures, where
bound nuclei can exist, interesting phases of nuclear matter are seen.
Temperatures as low as a few MeV are therefore outside the scope of these
discussions. In this discussion we have also assumed that at $\mub^1$
the baryon density is not high enough for colour superconducting (CS)
phases to manifest, an assertion which must be tested by other means.

\section{The $\muq$-$T$ plane}\label{sec:three}

The electric charge of a system is a conserved quantity which is
easy to measure in experiments, so $\muq$ is an useful quantity to
consider. However, most of the literature instead uses the chemical
potential, $\mui$, connected to the isospin projection $I_3$. Fortunately
the Gell-Mann-Nishijima relation $I_3=Q-(B+S)/2$ allows us to relate
the two. It shows that changes in $Q$ and $I_3$ are equal at fixed $B$
and $S$. Then equating the energy change due to such fluctuations in
the two ensembles implies that $\mui=\muq$.

When the u and d flavours of quarks are degenerate in mass, then an
isospin chemical potential breaks the SU(2) vector symmetry remaining
after chiral symmetry breaking into the U(1) symmetry generated by
$\tau_3$. A computation in a chiral model showed that a pion condensate,
$\Pc$, forms at $\mui=m_\pi/2$ \cite{Son:2000xc}. The model computation
showed that the phase transition from the normal hadron to this pion
condensed BEC (\pic) phase is of second order, and that there is a
critical line that emerges from it. Since the remnant symmetry is U(1)
$\simeq$ O(2), the critical indices are expected to be in the same
universality class as that of O(2) Heisenberg ferromagnetism.

\begin{figure}
\centering
\includegraphics{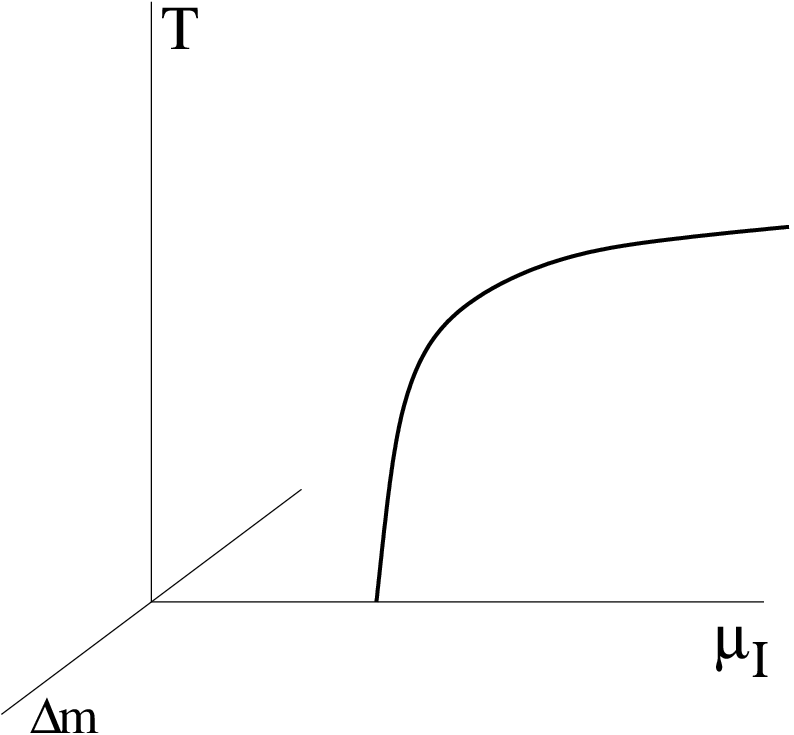}
\caption{The phase diagram for $N_f=1+1+1$ QCD in the space of $T$ and
$\mui$ as $\Delta m$ is varied. For $\Delta m=0$ lattice computations
indicate that the critical line between the hadron and {\pic} phases is
nearly vertical, until the cross over between the hadron and quark phases
is reached. Beyond this the critical line is either horizontal or rises
slowly. For any finite $\Delta m$ the transition turns into a crossover.
Since this whole volume has an extended symmetry, an isolated critical
line is allowed; this is the edge of a first order surface for a hidden
parameter, as explained in the text.}
\label{fg:muit}
\end{figure}

$N_f=2+1$ QCD computations on the lattice with $m_\pi \simeq135$ MeV
\cite{Brandt:2017oyy}, show a line of phase transitions starting at
$\mui=m_\pi/2$. The line is vertical within statistical errors until
it meets the line of cross overs between the hadron and quark phases,
and is subsequently nearly horizontal. This change of slope occurs at
$T\simeq155$ MeV, but this temperature is uncertain by about 5 MeV.
All this is consistent with the quantities reported in \eqn{curv}
and \eqn{bors}. A limited finite volume scaling has been performed to
check that the transition line is consistent with the expected O(2)
critical transitions. In a follow-up computation \cite{Brandt:2022hwy}
it was shown that the speed of sound $c_s>1/\sqrt3$, deep in the {\pic}
phase of pionic matter.

An interesting aspect of these lattice simulations is that a
``twisted mass'' is introduced in the light quark sector to stabilize the
flavour of the pion condensate \cite{Kogut:2002zg}. The Dirac operator
commutes with the flavoured transformation $\tau_2\gamma_5$ even when
the symmetry is broken in this way, so that the determinant remains free
of the sign problem. At finite $\lambda$ the coset space is ``tilted'' to
produce a preferred direction for the condensate, so there is no Goldstone
mode. This twisted mass parameter, $\lambda$, must be extrapolated to
zero to obtain the correct physics of the Goldstone boson in the chiral
limit of $N_f=1+1+1$.  However, Goldstone physics appears as vanishing
singular values of the Dirac operator which develop as $\lambda\to0$.
Even for small non-zero $\lambda$ they lead to a large condition number
and critical slowing down of the simulation. In \cite{Brandt:2017oyy}
the problem is partly resolved by treating the developing zero-modes
separately. Using this, a clever separation of the contribution of some
of the developing zero and the rest of the modes is used to define a
better behaved operator and partially circumvent the problem with the
extrapolation to $\lambda=0$. Several of the techniques used have matured
only in the last decade, enabling modern computations to overtake the
state of the art in \cite{Kogut:2002zg}.

Instead of $N_f=2+1$ if one examines the physical theory with $N_f=1+1+1$,
then $\Delta m$ also breaks SU(2) of isospin to the U(1) generated
by $\tau_3$.  Since exactly the same symmetry is broken by $\mui$,
switching this on in QCD with non-vanishing $\Delta m$ does not break an
existing symmetry, and so cannot produce a critical point. This implies
that there is only this line of second order phase transitions in the
phase diagram of physical QCD.  This is not consistent with Gibbs'
phase rule. However, physical QCD still has an unbroken symmetry since
there remains one massless mode. The parameter $\lambda$ breaks this
symmetry. The critical line at $\lambda=0$ is the boundary of a first
order region that arises for generic values of these parameters when
$\Delta m$ and $\mui$ are tuned appropriately. Although this parameter
vanishes in QCD, it must be considered as one of the theory parameters
of the phase diagram, on the same footing as $m$ and $\Delta m$ as far
as the thermodynamics is concerned.

In this way we understand that the phase diagram of QCD in the $\Delta
m$-$\mui$-$T$ space, shown in \fgn{muit}, is a space with an unbroken
symmetry, and hence allows the existence of a critical line without
a visible phase coexistence region. Interestingly, when the vector
symmetry is broken by $\Delta m$. then there is a sign problem for
lattice simulations at finite $\mui$.  Making a Taylor expansion of
the free energy \cite{Gavai:2003mf, Gavai:2004sd}, but now in $\mui$,
is an available method that is yet to be applied to this case.

Near any second order transition one can resolve $\Omega$ into
a singular and a regular piece. The contribution to the specific heat,
$\cv$, considering both pieces near the critical temperature $T_c$,
when $\mui$ is also tuned to criticality, gives
\begin{equation}
 \cv = A + \frac B{T_c|T-T_c|^\alpha},
\label{another}
\end{equation}
where $A$ comes from the regular piece and $B$ and $\alpha$ come from
the universal singular piece \cite{Gupta:2015dra}. Notice that $B$
can be negative as long as $\cv$ is positive. In fact, microgravity
experiments with liquid He \cite{Lipa:1996zz} show that this is so at
the $\lambda$-point of He, where the critical behaviour is in the O(2)
universality class. Since the universal quantities found in liquid He
are $\alpha\simeq-0.012$ and $B<0$, one may expect the same behaviour in
QCD at finite $\mui$. This means that $\cv$ is dominated by the regular
part of $\Omega$, and it would be hard to see a crossover using thermal
properties either in heavy-ion collisions or in neutron star interiors.

A recent lattice computation performed for finite $\mui$ and $T$
close to zero, for the first time found non-perturbative evidence for
a colour superconducting (CS) gap, $\Delta$, at the quark Fermi surface
for $\mui\ge1500$ MeV \cite{Abbott:2024vhj}. The computation used clover
improved Wilson quarks and an improved gauge action with pions which were
not too far from their physical mass. This study also brought together
several techniques which have matured only in recent years, and enabled
a first look at this exciting and long open problem.  The evidence for
CS came in the form of a disagreement between the pressure estimated
on the lattice and a gapless NNLO weak-coupling estimate. In the range
1500--3000 MeV, $\Delta$ was found to be roughly 300 MeV or lower (at
the upper end of the range $\Delta$ was consistent with zero). In this
range of $\mui$ a perturbative estimation of $\Delta$ \cite{Son:2000xc}
is consistent with the lattice result. There is a systematic trend which
needs to be investigated further: the perturbative estimate of $\Delta$
increases with $\mui$ whereas the lattice estimate is either constant
or decreasing. A follow up is also needed to check whether $\Delta$
remains non-zero till $\mui=\mui^1\simeq1000$ MeV.

\section{The phase diagram of physical QCD}\label{sec:four}

\begin{figure}
\centering
\includegraphics[scale=0.4]{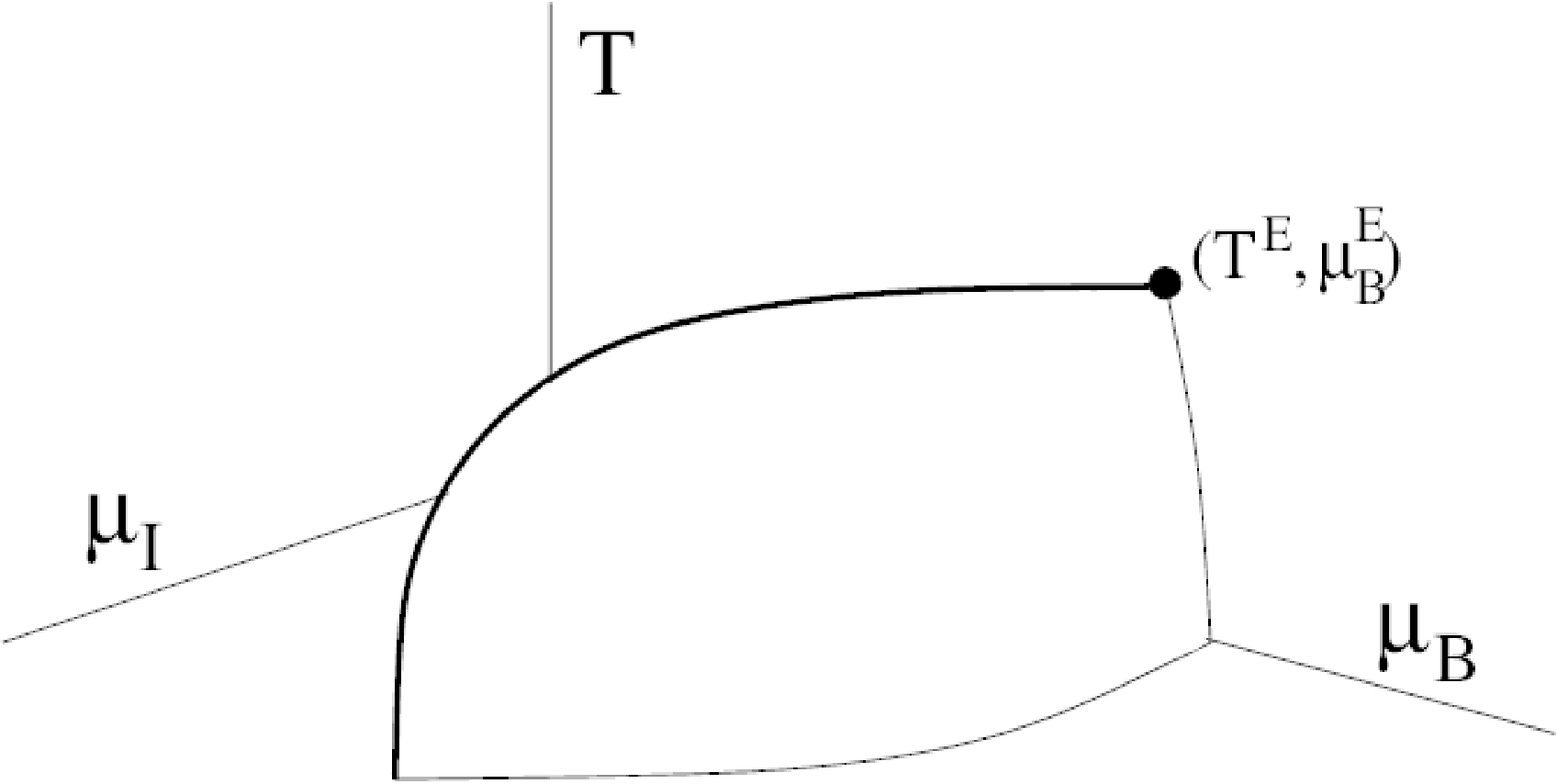}
\caption{The conjectured phase diagram of physical QCD for $\mus=0$,
based on currently available results from lattice QCD. There is a phase
coexistence surface separating the hadron and quark phases which ends
along a critical line. This curves down to meet the $T=0$ plane at the
cold critical point (CCP). This accommodates the strong possibility that
there is a continuity between the hadron and colour superconducting phases.}
\label{fg:phys}
\end{figure}

Enough information is available now from lattice simulations to
patch together the full phase diagram of physical QCD, \ie, QCD
with $N_f=1+1+1$. Various caveats about the pieces which go into
this have been mentioned already, but we bring this list together
again. At very low temperatures and $\mub$ between about 900 MeV
and $\mub^1$, there are likely to be interesting phases of nuclear
matter. These range from superconducting phases due to pairing forces
between nucleons \cite{Sedrakian:2006mq} and crystalline structure
\cite{Glendenning:2001pe} at the low-density end to nuclear pasta phases
\cite{Watanabe:2000rj, Maruyama:2005vb, Avancini:2008kg} towards higher
densities. There is a gap in the literature regarding the structure of
these phases at high $\mui$. This is clearly an area ripe
for quick exploration. At temperatures below about 100 MeV and around
$\mub^1$ there is a possibility of encountering colour superconducting
(CS) phases \cite{Berges:1998rc, Rajagopal:2000wf, Alford:2001zr}.
If they exist at these $\mub$ and $\mui$, they would definitely affect
the boundaries of the quark phase. In view of the current uncertainties
regarding the boundaries of these interesting phases, we first present
a construction that ignores them, before speculating on how they can
affect the tentative conclusions that could be obtained in this way.

We have considered two order parameters: one is the quark condensate
$\Sc$ which distinguishes the hadron and the quark phases, the other is
$\Pc$, the pion condensate which distinguishes the hadron and {\pic}
phases. Current thinking is that that in the plane $T$-$\mub$ there
is a first order line with phases distinguished by $\Sc$.  By Gibbs'
phase rule, this must develop into a surface in the phase diagram
of $T$-$\mub$-$\mui$. An alternative is the parsimonious scenario in which
there is no first order phase transition for physical QCD in this plane.
Heavy-ion collisions might be able to distinguish between these scenarios
unless, unluckily, the critical point exists but at too large a value of
$\mub/T$ to show up in experiments.  In the $T$-$\mui$ plane for physical
QCD there is either a mild crossover between the hadron and {\pic} phases,
or no crossover at all. We have argued that the crossover, if it exists,
may be hard to detect.

\begin{figure}
\centering
\includegraphics[scale=0.7]{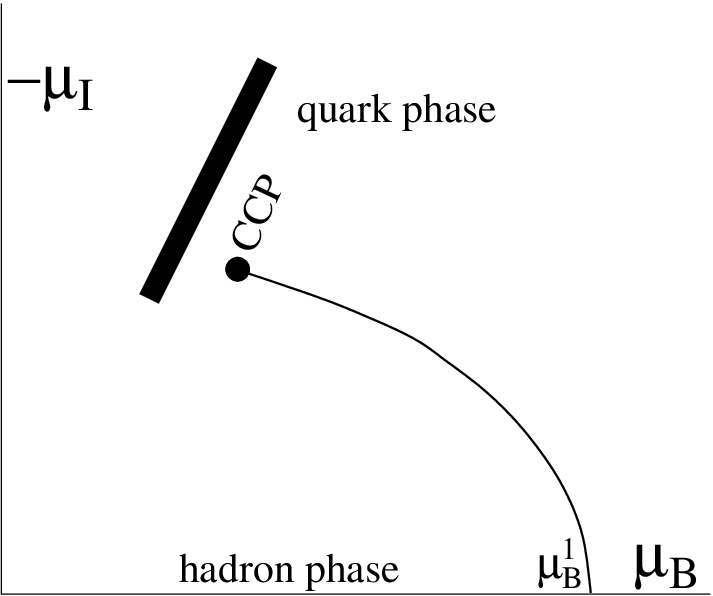}
\caption{The phase diagram for NSs at low $T$ which follows from
\fgn{phys}. The possible equilibrium states that cores of NSs can explore
is shown by the diagonal box. The upper end of the box corresponds to a
little over twice the normal nuclear density, and the lower end is the
normal nuclear density. We assumed that the CCP lies at larger $\mub$
than is explored by NSs so that the transition between hadron and quark
matter in NS cores is continuous and does not involve a latent heat. This
assumption remains to be tested.}
\label{fg:neust}
\end{figure}

The hadron-quark (HQ) coexistence surface touches the $\mub$ axis
at $\mub^1$. The sign of the curvature $\kappa_2^\Q$ in \eqn{curv}
indicates that if this were the only coexistence surface in the phase
diagram, then it would meet the $\mui$ axis at about 900 MeV. So the HQ
surface must bend towards the $\mui$ axis.  The question of whether the
hadron-quark coexistence surface meets the $\mui$ axis is interesting.
Some arguments are given in \cite{Son:2000xc} that {\pic} goes into
CS quark matter without a phase transition in $N_f=2+1$ QCD. This
is consistent with the observations in \cite{Abbott:2024vhj}.  So the
balance of evidence now is that the HQ coexistence surface ends before it
can meet the $\mub=0$ plane, this time at large $\mui$. This means that
there is a critical point in the $\mub$-$\mui$ plane for physical QCD at
$T=0$, \ie, a cold critical point (CCP).  A resulting parsimonious phase
diagram is that the Ising line in \fgn{phys} bends down to meet the $T=0$
plane, and is the only boundary of the HQ coexistence surface. In this
case the CCP is in the Ising universality class. In
any case, hadron-quark continuity along the $\mui$ axis needs to be
investigated further. A start can be made on the lattice with $\Delta
m=0$. An alternative to the scenario in \fgn{phys} is that there are no
phase transitions at all. This is not yet ruled out.

There are unresolved questions still about the physics when both $\mub$
and $\mui$ are non-vanishing.  For example, in \cite{Brandt:2022hwy} it
was found that for a large range of $\mui$ the speed of sound exceeds
the conformal limit.  Does this still happen when $\mub>0$? And with
non-vanishing $\Delta m$? Does the speed of sound drop in the vicinity of
the O(2) critical line, \ie, at small $\Delta m$? Lattice computations
will be hard with two independent sign problems, but EFTs should help
to clarify the range of possibilities.

Heavy ion collisions start with nuclei which are reasonably close
to isoscalar ($I_3/A=Z/A-1/2\simeq0.1$ for both Pb$^{82}_{208}$ and
Au$^{79}_{197}$), and come closer to isoscalarity in the central region of
the fireball. So the equilibrium states reached in these experiments lie
on a plane close to vanishing $\mui$ parallel to the $T$-$\mub$ coordinate
plane. They could either give evidence to support the existence of the
HQ coexistence surface, or give limits on the values of $\mub/T$ where it
can exist.  The cores of NSs, on the other hand, are as far from isoscalar
as it is possible for a stable system to go. Newly formed NSs have a
temperature of about 10 MeV, and they cool to around 100 KeV in about a
million years. So at any given age, NS matter explores a plane of small,
but non-vanishing, $T$ parallel to the $\mub$-$\mui$ coordinate plane.

The phase diagram in the plane explored by the cores of NSs is shown
in \fgn{neust}. The coexistence line between hadron (\pic) and quark
(perhaps CS) phases could end in a CCP (which we argued before is
possibly in the Ising universality class).  Since the baryon density in
a NS is almost completely due to neutrons, $\mui\simeq-2\mub$. Neutron
star cores lie along this line with $\mub$ corresponding to somewhere
between nuclear saturation density and twice that.  Current lattice
studies then indicate that the cores are either in the hadron/{\pic}
phase or in a quark phase. Whether this quark phase is CS is not yet
determined by lattice QCD. Note that the track followed by the NS core
may or may not cross a first order transition depending on the as yet
unknown position of the CCP. There is also a chance that there is no HQ
coexistence line, and the whole phase diagram is trivial.

In binary NS collisions, when matter is initially heated, it goes out of
thermal equilibrium. However, if the result of the merger comes back to
thermal equilibrium, then it does so at a higher temperature. Possibly
this could be a significantly higher temperature than of order 10 MeV.
Also, since the merged object is heavy, it would be at a higher density.
As a result, the remnant, whether it eventually becomes a NS or a
black hole, could pass through a phase of quark matter, possibly colour
superconducting quark matter. Unfortunately, this object is hidden inside
a hot cloud of radioactive nuclei synthesized during the collision.
Penetrating signals from this merged object are therefore interesting
to search for.

It is exciting that in the last ten years or so, lattice studies have
begun to constrain the possibilities for the physics at the core of NSs.
Many of the conjectures first made about 25 years ago are being tested,
and, as discussed here, this leads to new conjectures for the phase
diagram of QCD.  We have discussed at various places in this review
where lattice, EFT, or model studies can lead to greater constraints
on the possible physics. This is an exciting time in the physics of hot
and dense strongly interesting matter.

\section{Chiral transition and neutron stars at large $N_c$}\label{sec:five}

Some aspects of strong interactions are simpler to examine in the
't Hooft limit, $N_c\to\infty$ while holding $g_s^2N_c$ fixed. It was
argued by 't Hooft \cite{tHooft:1973alw}. that in this limit one can use
weak-coupling arguments to understand aspects of QCD which are otherwise
nonperturbative. In this section we apply this counting to aspects of
the QCD chiral transition and to neutron stars.  Since the physics of
neutron stars involves charge neutrality and stability against weak
decays, they cannot be understood within the strong interactions only.
The electroweak interactions play an important role. If one wants to take
the limit of the number of colours, $N_c$, to be large while retaining
realistic physics, one has to consider the limit in the appropriate
extension of the standard model.

We will take the minimal particle content needed, which is two flavours
of quarks, and the electron and the neutrino.  The chiral symmetry of
QCD is then SU$_L$(2) $\times$ SU$_R$(2) $\times$ U$_B$(1). The gauge
group is SU($N_c$) $\times$ SU$_L$(2) $\times$ U$_Y$(1) where $Y$ is the
weak hypercharge and we have the usual chiral form of the weak interactions.
With just one generation of quarks, global SU(2) anomaly cancellation
requires odd $N_c$.  Then with the usual assignment $Y=\pm N_c+1$ to the
u and d quarks and unity for leptons, we find first,
\begin{equation}
  Q_u-Q_d=1, \qquad{\rm and\ second,}\qquad Q_u^2-Q_d^2=\frac1{N_c^2}
\label{charge}
\end{equation}
from ABJ anomaly cancellation \cite{Chow:1995by,Shrock:1995bp}. We will
take the lepton charges to be as usual. However, the mixing amplitude
for $\rho$ and $\gamma$ varies as $ef_\rho$, where the $\rho$ decay
constant $f_\rho\propto\sqrt{N_c}$. In order to keep the amplitude
from growing, and hence pushing up the $\rho$ mass, the 't Hooft limit
requires the scaling $e^2N_c\to$ constant as $N_c$ is taken to infinity
\cite{Chow:1995by}. Similar arguments as $N_c$ is taken to infinity for
internal $W^\pm$ and $Z$ lines require the scaling $g_2^2N_c$ to go to
a constant, where $g_2$ is the electroweak coupling. The only remaining
parameters in the Lagrangian are the fermion masses, $m$, and $\theta_W$,
which are both taken to be fixed in the 't Hooft limit. With this,
the scalings discussed by 't Hooft \cite{tHooft:1973alw} and Witten
\cite{Witten:1979kh} can be recovered.

The flavour multiplets of the mesons are the same as for $N_c=3$. However,
since we build baryons by taking $N_c$ quarks, they can have isospins
between $1/2$ and $N_c/2$. In the doublet representation one has the large
$N_c$ analogue of the proton with $(N_c+1)/2$ up quarks and $(N_c-1)/2$
down quarks, so that its charge is unity. The isospin flipped state has
vanishing charge, and is the large $N_c$ analogue of the neutron. As
long as the $u$ quark is lighter than the $d$ quark, one can have the
$\beta$ decay $n\to pe^-\overline\nu$. In the large $N_c$ limit the
tower of states with equal spin and isospin ($J=I=1/2,3/2,5/2,\cdots$)
are degenerate, giving an effective SU(4) spin-flavour symmetry which
is similar to the quark model \cite{Gervais:1983wq,Gervais:1984rc}.

To proceed one needs to examine these baryon masses in QCD. The
small current quark mass which appears in the Lagrangian of the
Standard Model is often supplemented with a constituent quark mass,
$\mcons$, which is such that the $\rho$ meson mass is about $2\mcons$
(since quark and antiquark masses are equal), and the nucleon mass
is about $3\mcons$. Generalizing this using Witten's counting rule
\cite{Witten:1979kh} would give us $\mcons\propto N_c^0$. In models
of chiral symmetry breaking the constituent masses of the quarks
arise from the chiral condensate, $\ppbar$. Since $m_\pi^2 f_\pi^2 =
m\ppbar$, with $m_\pi\propto N_c^0$ and $f_\pi\propto N_c^{1/2}$, one
has $\ppbar\propto N_c$ \cite{Bonanno:2025hzr}. In such models one also
finds $\mcons\simeq\ppbar/N_c\propto N_c^0$, which is consistent with
the above.  This argument implies that quark masses are fixed in some
units independent of $N_c$, and that baryon masses increase when $N_c$
increases. Furthermore, the mass degeneracy that gives SU(4) spin-flavour
symmetry is lifted at order $1/N_c$ \cite{Jenkins:1993zu}. For the baryon
masses one finds

\begin{equation}
  m_\B = N_c \left[a_0 + a_1 \left(\frac J{N_c}\right)^2 
     + a_2 \left(\frac J{N_c}\right)^4 + \cdots \right],
\label{mbaryon}
\end{equation}
with constants $a_i$ to be determined and with the $J=1/2$ doublet having
the lowest mass.

\begin{figure}
\centering
\includegraphics[scale=0.75]{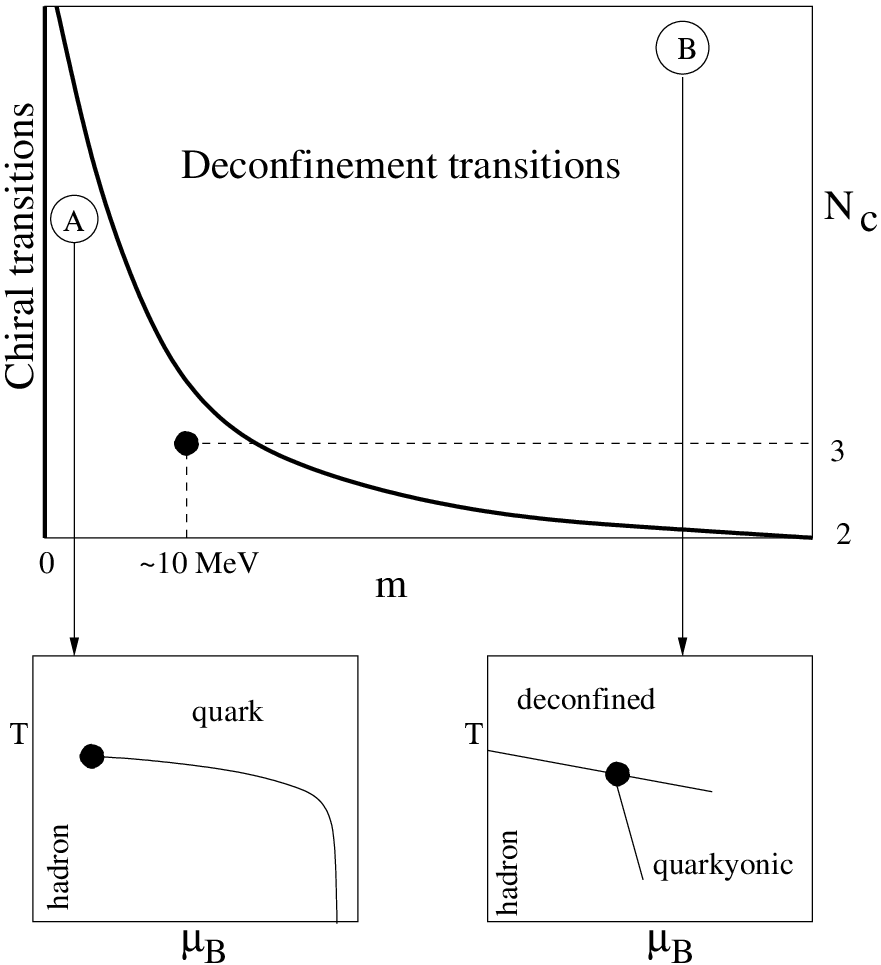}
\caption{The most parsimonious model of the relationship between the
chiral and deconfinement transitions with changing $N_c$ which follows
from our knowledge of $N_c=3$. The difference between $T_c$ and $T_d$
cannot be resolved through power counting alone. The regions labelled
A and B give different phase diagrams, as shown. The phase diagrams
that are discussed in the previous sections hold in region A.}
\label{fg:largen}
\end{figure}

This is a good place to discuss phase transitions at finite temperature.
In the chiral limit of QCD, the chiral condensate, $\ppbar$, is an
order parameter for the chiral symmetry restoring phase transition. The
phase diagrams of the previous sections have been constructed using
this order parameter and related physics. The chiral transition
temperature, $T_c$ in \eqn{chiral}, is expected to scale as $N_c^0$.
When the lightest quarks are very massive, then one expects QCD to
have a deconfining phase transition, with an order parameter called
the Polyakov loop. These considerations are incorporated into the
Columbia plot \cite{Brown:1990ev}, and a deconfining transition
temperature $T_d=192\pm4\pm7$ MeV was reported in \cite{Cheng:2006qk}.
In \cite{McLerran:2007qj} it is argued that that $T_d$ is expected to
scale as $N_c^0$. Scaling arguments cannot resolve whether $T_c<T_d$
in the 't Hooft limit. A parsimonious argument about the relationship
between these phase transitions was presented in \cite{Datta:2009jn}
and summarized in \fgn{largen}.

Within the EFT discussed in Section \ref{sec:two} it is possible
to construct a large-$N_c$ argument which shows that the curvature
coefficient $\kappa_2$ is two powers of $N_c$ smaller than $T_c$ and that
$\kappa_{n+2}$ is two powers of $N_c$ down from $\kappa_n$. As a result,
in the large $N_c$ limit $T_c(\mub)$ becomes independent of $\mub$. It
is interesting to note that both the chiral critical line (in region A of
\fgn{largen}) and the first order deconfinement transition line (in region
B of \fgn{largen}) are independent of $\mub$ in the 't Hooft limit. It
is also interesting to note that since the baryon mass is of order $N_c$
whereas meson masses are of order $N_c^0$, baryonless thermal EFTs may
continue to play an important role in the chiral symmetric phase of QCD.

Returning next to the physics of baryons, we see that baryon-baryon
interactions which give rise to nuclear physics, as well as interactions
of baryons with electrons or photons are subleading in $N_c$. So, one
might find that models of stellar physics simplify in the large $N_c$
limit. However, one has to take the 't Hooft limit in such a way as to
ensure that not all stars become black holes.  The mass of a star with
$A$ nucleons, say the same as the baryon number of the sun, scales as
$\msol\propto N_c$.  The Schwarzschild radius of a black hole with this
mass scales as $N_c/\gnew$. In order to leave space for normal stars,
this cannot go to infinity as $N_c\to\infty$. This requires that to
take the 't Hooft limit one should scale down the Newton constant by
taking $\gnew N_c$ to be a constant. By taking the constant to be zero,
one can also accommodate a faster decrease of $\gnew$ with $N_c$. Note
that the large-N limit used in the AdS/CFT correspondence requires a
different scaling $\gnew N_c^2$ being held constant. This faster fall
of Newton's constant is also compatible with a separation between the
radii of black holes and stars in the 't Hooft limit.

The $I=J=1/2$ flavour multiplet contains the large $N_c$ analogues of the
neutron and proton. As long as the temperature of matter lies below the
splitting between the baryon doublet and higher multiplets, \ie, for $T$
less than of order $1/N_c$, one can consider matter to consist of just
the lowest doublet of baryons and the electron.  Local charge neutrality
and stability against $\beta$-decay and electron capture of cold matter
then requires two conditions on the Fermi energies \begin{equation}
  m_p E_F^p = m_e E_F^e, \qquad{\rm and}\qquad E_F^n = E_F^p+E_F^e,
\label{efermi} \end{equation} provided that the decay neutrinos escape
from matter. Then writing $\epsilon = m_e/m_p\propto1/N_c$, one finds
that $E_F^p = \epsilon E_F^e$ and that $E_F^n\simeq E_F^e$. If the
star has nucleon number $A$, then one finds the proton number to be
$Z\propto A/N_c^{3/2}$. In the limit a neutron star can consist of
only neutrons and electrons, the latter also being neutral due to the
scaling of the electron charge in the 't Hooft limit. Interestingly,
magnetic field effects can be interesting in this limit only if magnetic
field strengths are of order $\sqrt{N_c}$.  BNS mergers which raise the
temperature of matter by more than order $1/N_c$ would then excite the
full ladder of quark model baryons and possibly cross over into a state
with quark matter.

\section{Acknowledgements}\label{sec:six}

I would like to thank Bastian Brandt, Heng-Tong Ding and Gergeli Endr\"odi
for useful discussions and critical readings of the manuscript.

\end{document}